# Statistical mechanics of a nonequilibrium steady-state classical particle system driven by a constant external force

Jie Yao （姚婕）[1,2] and Yanting Wang（王延颋）[1,2,*]

[1]*CAS Key Laboratory of Theoretical Physics, Institute of Theoretical Physics, Chinese Academy of Sciences, 55 East Zhongguancun Road, P. O. Box 2735, Beijing, 100190 China*

[2]*School of Physical Sciences, University of Chinese Academy of Sciences, 19A Yuquan Road, Beijing 100049, China*

**Abstract.** A classical particle system coupled with a thermostat driven by an external constant force reaches its steady state when the ensemble-averaged drift velocity does not vary with time. The statistical mechanics of such a system is derived merely based on the equal probability and ergodicity principles, free from any conclusions drawn on equilibrium statistical mechanics or local equilibrium hypothesis. The momentum space distribution is determined by a random walk argument, and the position space distribution is determined by employing the equal probability and ergodicity principles. The expressions for energy, entropy, free energy, and pressures are then deduced, and the relation among external force, drift velocity, and temperature is also established. Moreover, the relaxation towards its equilibrium is found to be an exponentially decaying process obeying the minimum entropy production theorem.

[*]Corresponding author: wangyt@itp.ac.cn.

# 1. Introduction

The theoretical framework for equilibrium statistical mechanics is well established, but that for non-equilibrium, except the linear response regime, is still in its early stage. Until now, the fluctuation theorems[1-6], represented by the Jarzynski equality[6], are the only set of equalities that hold for nonequilibrium systems arbitrarily far from equilibrium. The lack of theoretical frameworks for nonequilibrium statistical mechanics hinders the investigations to real systems since most of the time they work at non-equilibrium states. A simpler case is non-equilibrium steady states (NESS) whose thermodynamic properties do not change with time while a certain flux exists. For a NESS, a generalized force constantly input energy into the system, which is exactly counter-balanced by a certain dissipation process, and thus the ensemble-averaged flux does not vary with time. An applied NESS system can be the electrolyte in a battery driven by the electrostatic force between two electrodes (see, e.g., Refs. 7, 8). Surprisingly, there are no general theories even for the case as simple as NESS. A direct consequence is that molecular dynamics simulations of such kind of systems still have to approximately use the equilibrium thermostats, such as the Berendsen thermostat[9] and the Nosé-Hoover thermostat[10, 11], despite the fact that it is known to introduce, sometimes significant, errors into simulation results[12].

In this paper, we establish the statistical mechanics for a classical particle system driven by a constant external force and coupled with a constant-temperature thermostat, whose NESS is arbitrarily far from equilibrium. None of the conclusions drawn on equilibrium systems or non-equilibrium systems with local equilibrium hypothesis are adopted. That is, similar to the equilibrium case, the deduction is solely based on the equal probability and ergodicity principles. As shown in Figure 1, the system contains $N$ identical classical particles coupled to a thermostat with a constant temperature $T$ driven by a constant force $F$ along the $x$-axis. After the system reaches the steady state, the macroscopic drift velocity $V_d$ along the $x$-axis is a constant. Because the system is still canonical (the external potential is stationary), the momentum and position spaces are independent of each other.

In the following sections, we will first determine the momentum space and position space distributions, followed by the calculations of system energy, entropy, free energy, and pressures. The relation among the external force $F$, the drift velocity $V_d$, and the temperature $T$ is then established. Finally, the relaxation process towards its equilibrium is revealed to be an exponential decay obeying the minimum entropy production theorem.

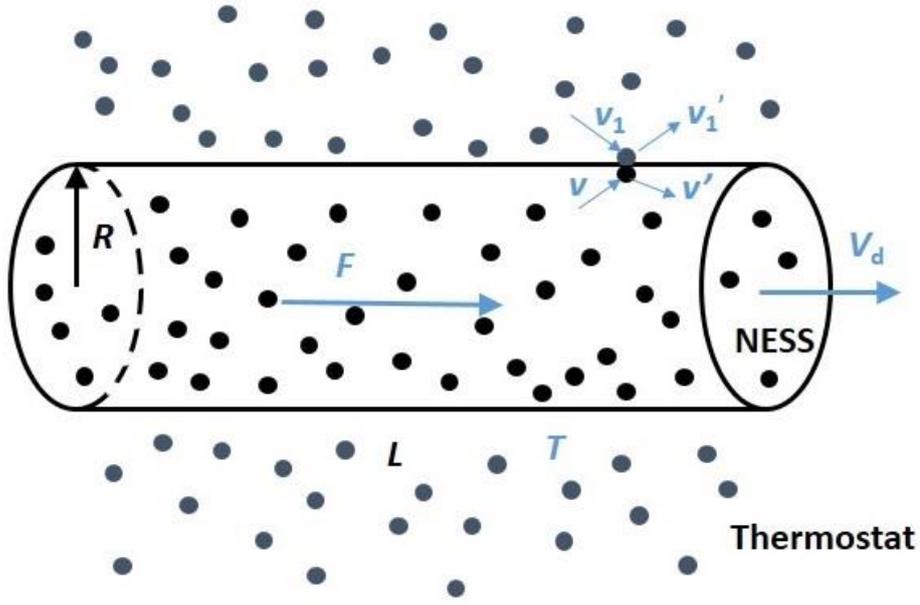

**Figure 1.** The schematic of the studied classical particle system in the NESS with *N* particles driven by a constant force *F* and coupled to a thermostat with a constant temperature *T*, resulting in a constant drift velocity $V_d$.

## 2. Momentum space distributions

Since for a canonical classical system, the momentum space is independent of the position space, the first question to ask is what the distributions in the momentum space, i.e., the velocity distributions are. Since the force is applied along the *x*-axis, obviously the velocities along the *y* and *z* directions $v_y$ and $v_z$ should both still obey the equilibrium Maxwell-Boltzmann distribution

$$f(v_y)dv_y = \sqrt{\frac{m}{2\pi k_B T}} \exp\left(-\frac{mv_y^2}{2k_B T}\right) dv_y,$$
$$f(v_z)dv_z = \sqrt{\frac{m}{2\pi k_B T}} \exp\left(-\frac{mv_z^2}{2k_B T}\right) dv_z,$$
(1)

where $k_B$ is the Boltzmann constant and *m* is the particle mass. For a random velocity along the *x* direction $v_x$, we assume that it is generated by *n*-times random walks with the magnitude of $\delta$ for each walk, and the direction can be either positive or negative. If the probability for taking the positive direction is *p*, then the probability of having *k* times positive walks in all *n* walks is described by the binomial distribution

$$P(k|n) = C_n^k p^k (1-p)^{n-k}. \tag{2}$$

When $n \to \infty$, the above approaches the Gaussian distribution

$$f(k)dk = \frac{1}{\sqrt{2\pi}\sigma} \exp\left(-\frac{(k-\mu)^2}{2\sigma^2}\right) dk, \tag{3}$$

where $\mu = np$ and $\sigma^2 = np(1-p)$. The corresponding velocity along the *x*-axis is $v_x = (2k-n)\delta$, which reaches its maximum value of $V_{max} = n\delta$ when $k = n$ and its minimum value of $-V_{max} = -n\delta$ when $k = 0$. Because $\langle k \rangle = \mu = np$, the thermodynamic drift velocity

$$V_d = \langle v_x \rangle = (2p-1)V_{max}. \tag{4}$$

Since $k = \frac{v_x}{2\delta} + \frac{n}{2}$, Eq. (3) becomes

$$f(v_x)dv_x = \frac{\sqrt{n}}{2V_{max}\sqrt{2\pi}\sqrt{p(1-p)}} \exp\left(-\frac{(v_x - (2p-1)V_{max})^2}{8p(1-p)V_{max}^2/n}\right) dv_x. \tag{5}$$

By requiring that both the expectation value and the variance of Eq. (5) should be constants when $n \to \infty$, we obtain $V_{max} \propto \sqrt{n}$ and $p - \frac{1}{2} \propto \frac{1}{\sqrt{n}}$, so

$$f(v_x)dv_x = \frac{\sqrt{n}}{V_{max}\sqrt{2\pi}} \exp\left(-\frac{(v_x - V_d)^2}{2V_{max}^2/n}\right) dv_x. \tag{6}$$

When $V_d = 0$, Eq. (6) should go back to the equilibrium Maxwell-Boltzmann distribution, so we obtain $\frac{V_{max}^2}{n} = \frac{k_B T}{m}$. The velocity distribution along the *x*-axis is thus

$$f(v_x)dv_x = \sqrt{\frac{m}{2\pi k_B T}} \exp\left(-\frac{m(v_x - V_d)^2}{2k_B T}\right) dv_x. \tag{7}$$

The same result can be reached by utilizing the local equilibrium hypothesis[13]. Nevertheless, our derivation indicates that this result is of a more general basis free of the local equilibrium hypothesis.

By utilizing the $\chi^2$-distribution, it is easy to show that the velocity perpendicular to be drift velocity $v_\perp = \sqrt{v_y^2 + v_z^2}$ obeys the probability distribution

$$f(v_\perp) = \frac{mv_\perp}{k_B T} \exp\left(-\frac{mv_\perp^2}{2k_B T}\right), \quad v_\perp \geq 0. \tag{8}$$

## 3. Position space distribution

The derivation of the position space distribution is analogous to the equilibrium case[14]. The energy of the NESS system is denoted as $U_t$, and the energy of the whole system composed of the thermostat and the NESS system is denoted as $U$. Both the whole system and the thermostat are treated as infinitely-large microcanonical ensemble systems, whose numbers of microstates are $\Omega(U)$ and $\Omega(U-U_t)$, respectively. The NESS system is also infinitely large, but infinitely small compared to the thermostat, so the energy dissipates to the thermostat is negligible. Given the entropy of a microcanonical system $S = k_B \ln \Omega$, the probability that the NESS system has the energy $U_t$ is

$$P(U_t) = \frac{1}{Z} \frac{\Omega(U - U_t)}{\Omega(U)} = \frac{1}{Z} \exp\left(\frac{S(U - U_t) - S(U)}{k_B}\right). \tag{9}$$

By utilizing the Taylor expansion and $T = \left(\frac{\partial U}{\partial S}\right)_{U_t = 0}$, we have

$$P(U_t) = \frac{1}{Z} \exp(-\beta U_t), \tag{10}$$

where the partition function

$$Z = \frac{1}{N! h^{3N}} \int d\mathbf{r}^N \int d\mathbf{p}^N \exp\left(-\beta U_t(\mathbf{r}^N, \mathbf{p}^N)\right), \tag{11}$$

in which $h$ is the Plank constant, $\mathbf{r}^N$ and $\mathbf{p}^N$ are the collections of particle positions and momenta, respectively. For a classical canonical system, the momentum space is independent of the position space, so the partition function $Z = QZ_{id}$, where $Q = \int d\mathbf{r}^N \exp(-\beta U_p(\mathbf{r}^N))$ is the configuration integral, and $Z_{id}$ is the partition function of the ideal gas NESS system. Because $v_x \in (-\infty, +\infty)$, according to Eq. (7), the expression of $Z_{id}$ is exactly as the partition function of the equilibrium ideal gas system.

Although temperature is well defined for equilibrium systems, currently its rigorous definition for nonequilibrium systems is still lack[15]. In this work, it should be emphasized that the temperature $T$ and associated relations are defined for the equilibrium microcanonical thermostat as well as the whole system, not directly defined for the NESS system.

## 4. Entropy and free energy

Because the system entropy expression $S = -k_B \sum_i P(U_i) \ln P(U_i)$ can be deduced solely based on the equal probability and ergodicity principles, it is also applicable to the NESS system. Because $v_x \in (-\infty, +\infty)$, according to Eq. (7), the expression of $P(U_i)$ is identical for both equilibrium and NESS systems, and thus their entropy is the same. According to $F = U - TS$, the free energy of a NESS system is $E_d \equiv \frac{N}{2} m V_d^2$ larger than its equilibrium counterpart, exactly as the total energy difference. In other words, in principle the drift energy $E_d$ can all be used to do external work.

## 5. Injection-dissipation relation

Next we determine the relation among the external force $F$, the drift velocity $V_d$, and the temperature $T$ according to the balance between energy injection and dissipation. During a time interval $\Delta t$, the system kinetic energy increment caused by $F$ is

$$\Delta E_K = N \int \frac{m}{2} \left[ \left( v_x + \frac{F}{m} \Delta t \right)^2 - v_x^2 \right] f(v_x) dv_x, \quad (12)$$

so the injection power

$$w = \lim_{\Delta t \to \infty} \frac{\Delta E_k}{\Delta t} = NFV_d. \quad (13)$$

On the other hand, the dissipation power can be calculated by considering the energy exchange between the system and the thermostat. First consider the energy exchange by perfect elastic collisions on the boarder of the cylindrical container along the perpendicular direction (see Figure 1). The energy change of a NESS particle in a single-time collision can be obtained as $\Delta u = \frac{m}{2}(\mathbf{v}_1^2 - \mathbf{v}^2)$, where $\mathbf{v}$ and $\mathbf{v}_1$ are the velocities of the NESS and thermostat particles, respectively. However, the input energy cannot be sufficiently dissipated if the system particles only exchange energy on the boarder, because the external force $F$ is applied to all system particles; when the system size goes to infinity, the input power grows with the volume but the dissipation power grows with the area. Therefore, we have to assume that all system particles can experience "virtual" collisions with thermostat in the whole system volume, and the collision times $n_c$ for each particle in a unit time interval is proportional to the particle velocity in the direction perpendicular to the force:

$$n_c(v_\perp) = c_0 v_\perp, \tag{14}$$

where $c_0$ is a coefficient reflecting the coupling strength between the system and the thermostat. Because the number of particles with a perpendicular velocity of $v_\perp$ is $Nf(v_\perp)$, the total collision times for all particles with the perpendicular velocity of $v_\perp$ in a time interval $\Delta t$ is

$$M_c(v_\perp) = Nf(v_\perp) n_c(v_\perp) \Delta t, \tag{15}$$

where the probability distribution of the perpendicular velocity is given by Eq. (8). Therefore, the dissipation power, which has the exact value of $w$ but an opposite sign, can be calculated as

$$-w = \int_0^\infty dv_\perp \frac{M(v_\perp)}{\Delta t} \int d\mathbf{v} f(\mathbf{v}) \int d\mathbf{v}_1 f(\mathbf{v}_1) \Delta u = -NC_0 \sqrt{T} V_d^2, \tag{16}$$

where $C_0 \equiv c_0 \sqrt{\frac{\pi m k_B}{8}}$ is a coefficient whose exact value may vary with the coupling strength between the NESS system and the thermostat. Therefore, the relation among $F$, $V_d$, and $T$ is

$$F = C_0 V_d \sqrt{T}. \tag{17}$$

## 6. Pressures

We then calculate the pressures of the ideal gas NESS system. For a more general case, the contribution from particle interactions can be added in the same way as for equilibrium systems. Under the perfect elastic collision assumption, the momentum change for a single collision of the NESS particle on the boarder is $\Delta p(v_\perp) = 2mv_\perp$. According to the ergodicity and equal probability principles, the ensemble-averaged particles density in the system space is uniform, so the times for particles colliding the container surface in $\Delta t$ is

$$M_s(v_\perp) = Nf(v_\perp)\frac{\Delta V(v_\perp)}{V} = Nf(v_\perp)\frac{\pi\left[R^2 - (R - v_\perp \Delta t^2)^2\right]L}{\pi R^2 L},$$
$$\approx Nf(v_\perp)\frac{2v_\perp \Delta t}{R} = \rho S v_\perp \Delta t f(v_\perp) \tag{18}$$

where the particle density $\rho \equiv \frac{N}{V}$, $R$ and $L$ are the radius and side length of the cylinder, respectively, and $S = 2\pi RL$ and $V = \pi R^2 L$ are the area and volume of the cylinder, respectively, as shown in Figure 1. Then the pressure perpendicular to the flux is

$$P_\perp = \int_0^\infty dv_\perp \frac{M_s(v_\perp)}{S\Delta t} \Delta p(v_\perp) = 4\rho k_B T, \tag{19}$$

Note that the above expression includes the pressure along four directions, namely $y^+$, $y^-$, $z^+$, and $z^-$. It satisfies the equilibrium ideal gas equation of state along each direction. Along the flux direction, the times for a particle colliding a cylindrical surface is $M(v_x)dv_x = \rho \pi R^2 \Delta t v_x f(v_x) dv_x$, and the momentum change for each collision is $\Delta p(v_x) = 2mv_x$, so the pressure along the $x^+$ direction is

$$P_{x+} = \frac{1}{\pi R^2 \Delta t}\int_0^\infty dv_x M(v_x) 2mv_x$$
$$= 2m\rho \int_0^\infty v_x^2 \sqrt{\frac{m}{2\pi k_B T}} \exp\left(-\frac{m(v_x - V_d)^2}{2k_B T}\right) dv_x \tag{20}$$

Similarly, the pressure along the $x^-$ direction is

$$P_{x-} = 2m\rho \int_{-\infty}^0 v_x^2 \sqrt{\frac{m}{2\pi k_B T}} \exp\left(-\frac{m(v_x - V_d)^2}{2k_B T}\right) dv_x. \tag{21}$$

The integrals in Eqs. (20) and (21) do not have analytical solutions and can be solved numerically.

## 7. Relaxation towards equilibrium

If the constant external force $F$ is applied to the system when time $t<0$ and suddenly removed at $t=0$, the system will relax from NESS towards its equilibrium state by dissipating the drift energy $\frac{N}{2}mV_d^2(t=0)$ into the thermostat gradually. During an infinitesimal time interval $dt$, the energy change equals to the energy dissipation:

$$\frac{d}{dt}\left(\frac{N}{2}mV_d^2(t)\right) = -w = -NC_0\sqrt{T}V_d^2(t). \tag{22}$$

The solution of the above equation is

$$V_d(t) = V_d(0)\exp\left(-\frac{C_0\sqrt{T}}{m}t\right), \tag{23}$$

where $V_d(0)$ is given by Eq. (17).

As we have shown in section 4, for an NESS with an arbitrary $V_d(t)$, its entropy equals to the corresponding equilibrium system ($V_d = 0$). Therefore, we always have

$$\frac{dS(t)}{dt} = \frac{dS_{exch}(t)}{dt} + \frac{dS_{int}(t)}{dt} = 0 \tag{24}$$

at any time $t$ during the decaying process, where $dS_{exch}(t)$ is the entropy change flowing from the system to the thermostat, $dS_{int}(t)$ is the internal entropy change inside the system. According to Eq. (13), the dissipation heat from the NESS system to the thermostat in an infinitesimal time interval $dt$ is

$$dQ(t) = -w(t)dt = -NC_0\sqrt{T}V_d^2(t)dt. \tag{25}$$

On the other hand, we have

$$dQ(t) = TdS_{exch}(t), \tag{26}$$

so

$$\frac{dS_{int}(t)}{dt} = -\frac{dS_{exch}(t)}{dt} = \frac{NC_0}{\sqrt{T}}V_d^2(t). \tag{27}$$

Substituting Eq. (23) into the above equation, we obtain the entropy production

$$P_S(t) \equiv \frac{dS_{int}(t)}{dt} = \frac{NC_0}{\sqrt{T}}V_d^2(0)\exp\left(-2\frac{C_0\sqrt{T}}{m}t\right) \geq 0. \tag{28}$$

The change of the entropy production with time is thus

$$\frac{dP_S(t)}{dt} = -2N\frac{C_0^2}{m}V_d^2(0)\exp\left(-2\frac{C_0\sqrt{T}}{m}t\right) \leq 0. \tag{29}$$

The above two expressions indicate that the decaying process has its entropy production monotonically decreases and approaches its minimum value of 0 at $t \to \infty$ when the system approaches equilibrium, which agrees with the minimum entropy production theorem by I. Prigogine[16].

## 8. Conclusions

In this work, solely based on the ergodicity and equal probability principles, we derive the statistical mechanics of a NESS system coupled to a thermostat with temperature $T$ driven by a constant force $F$. A random-walk analysis determines that the momentum space distribution differs from its equilibrium counterpart only by a shift of the drift velocity $V_d$ in the distribution of the velocity along the force direction. The position space distribution is determined in the same way as for an equilibrium system in the canonical ensemble and is found to have exactly the same expression as the equilibrium counterpart. The entropy of the NESS system is found to be exactly the same as the corresponding equilibrium system, which means that the drift energy equals to the free energy difference and can all be used to do external work. The relation among $F$, $T$, and $V_d$ is established by considering the balance between the input power and dissipation power: $V_d$ is proportional to $F$ and $T^{-1/2}$. The pressure towards the force is larger than that along the directions perpendicular to the force, and the pressure along the force is smaller. Finally, the relaxation towards equilibrium is proved to be an exponential decay obeying the minimum entropy production theorem.

Although the studied NESS system is the simplest non-equilibrium thermodynamic system, it is remarkable that its statistical mechanics features far from equilibrium can be quantified without any approximations and only based on the ergodicity and equal probability principles. The results of this work not only allow the development of a thermostat algorithm for molecular dynamics simulations of NESS systems, but also serve as a starting point for further investigation of more complex non-equilibrium systems more common in nature than equilibrium systems, such as biological and social systems.

## Acknowledgments

This work was funded by the Strategic Priority Research Program of Chinese Academy of Sciences (Grant No. XDA17010504) and the National Natural Science Foundation of China (Nos. 11774357, 11947302).